\documentstyle[12pt]{article}
\begin{document}

\begin{center} Further Comments On The Effects of Deformation on Isovector \\
Electromagnetic and Weak Transition Strengths \medskip \\ 
Shadow J.Q. Robinson$^{a}$, 
L. Zamick$^{a,b}$, A. Mekjian$^{a}$,
N. Auerbach$^{a,c}$  
\end{center}

\noindent a) Department of Physics and Astronomy,
Rutgers University, Piscataway, \\New Jersey  08855

\noindent b)  TRIUMF,4004 Wesbrook Mall, Vancouver, British 
Columbia, \\Canada, V6T 2A3

\noindent c) Permanent address, School of Physics, Tel-Aviv
University, Tel Aviv, \\Israel.

\begin{abstract}
We present a superior proof that the results for summed strength isovector 
dipole, spin dipole, and orbital dipole excitations are independent of 
deformations at the $\Delta$ N = 0 level.  The effects of 
different oscillator frequencies in the 
x, y, and z directions are also considered.

\end{abstract}
\vspace{2.5in}

\newpage
\noindent 1)  INTRODUCTION
\bigskip

As has been previously noted [1], using the rotational model for $^{12}C$ and harmonic 
oscillator wave functions, the results for summed strength isovector dipole, 
spin-dipole, and orbital dipole excitations were independent of deformation 
in a $\Delta N = 0$ Nilsson model.  We then presented a more general proof 
which did not require the use of the rotational model explicitly but rather 
did require that the valence nucleons were all in the $0p$ shell and that 
the mean square radius of $p_{1/2}$ and $p_{3/2}$ particles were the same 
(as they are with harmonic oscillator wave functions).

We here present a superior proof and make several points about dipole excitations.  We 
consider excitations from the ground state of an $N=Z$ open shell nucleus (like 
$^{12}C$).  We will assume the ground state has angular momentum $J=0^{+}$.  As 
in the original work we consider the operators ($rY^{1}_{k}t$,
$r[Y^{1}s]^{\lambda}_{k}t$, and $r[Y^{1}\ell]^{\lambda}_{k}t$ .).  
Some of these operators arise in (p,n) reactions 
or neutrino absorption such as 
$\nu_{e} + ^{12}C \rightarrow ^{12}N + e^{-}$.

We show again in Table 1 the results of the summed strength in the asymptotic 
(oblate) limit and the spherical limit for the above operators in  $^{12}C$.  
The results to individual final momenta $\lambda$ are different in these two 
limits, but the total summed 
strength is the same in these two limits.

For the ordinary dipole operator $rY^{\prime}_{k}t_{+}$, the summed 
strength (SUM) , multiplied 
by $4\pi m \omega$/$\hbar$ is 27; for the spin dipole it is 20.25 and for the 
orbital dipole 48.  We will soon explain why this is so.
\bigskip

\noindent 2)  THE NEW APPROACH
\bigskip

To see why the results for SUM are independent of the specific $0p$ 
configuration (or deformation) when spherical harmonic oscillators 
(H.O.) wavefunctions are used we 
note the following unique feature of dipole excitations: In the  H.O. 
approximation
there is only one excitation energy, $1\hbar \omega$.  For the other 
modes this is not the case.  For $E2$ 
transitions, the strength of which are highly dependent on deformation 
there are both $0\hbar \omega$  and $2\hbar \omega$ excitations; for $E3$ 
we have $1\hbar \omega$ and $3\hbar \omega$ excitations etc.

Since for $E1$ transitions there is only \underline{one} excitation 
energy involved we can relate the summed strength to the energy 
weighted strength E.W.S.

\begin{equation}
SUM = E.W.S. / \hbar \omega
\end{equation}

The energy weighted strengths have been studied a great deal , and 
if we ignore, for the moment, the lack of commutivity of the potential 
energy with the various dipole operators, very simple results emerge.

Let us first show the electric dipole EWS referred to the center of mass,
as given in Bohr and Mottelson[2].  
They write the operator 
$M(E1,\mu) = e \sum_{i} (\frac{N-Z}{2A} - t_{3}(i))(rY_{\mu}^{1})_{i}$
The 'classical' EWS for this operator is 

\begin{equation}
EWS = \frac{9}{4\pi} \frac{\hbar^{2}}{2M} \frac{NZ}{A}
\end{equation}
Which for $N=Z=\frac{A}{2}$ becomes
\begin{equation}
EWS = \frac{9}{32\pi} A \frac{\hbar^{2}}{M}
\end{equation}

In our problem we have $t_{+}$ rather than $t_{z}$. Again going to the case of 
$N=Z=\frac{A}{2}$
, we have 
$M(E1,\mu) = e \sum_{i} (-t_{+}(i))(rY_{\mu}^{1})_{i}$
The EWS is now expressed as
\begin{equation}
\frac{1}{2}[EWS(+) + EWS(-)] = \frac{9}{8\pi}
\sum<0|[zt_{-},[-\frac{\hbar^{2}}{2M}\frac{d^{2}}{dz^{2}},zt_{+}]]|0>
\end{equation}
where EWS(+) is the energy weighted strength for a process in which 
a neutron is changed into a 
proton and EWS(-) where a proton is changed into a neutron.
Using the relations
\begin{equation}
[\frac{d^{2}}{dz^{2}},zt_{+}] = 2\frac{d}{dz}t_{+}
\end{equation}
\begin{equation}
[z,\frac{d}{dz}] = -1
\end{equation}
\begin{equation}
[t_{-},t_{+}] = -2t_{z}
\end{equation}
We are reduced to
\begin{equation}
\frac{1}{2}[EWS(+) + EWS(-)] = \frac{9}{8\pi} \frac{\hbar^{2}}{M} \sum 
<0| t_{z}+\frac{1}{2}+t_{z}2z\frac{d}{dz}|0>
\end{equation}
We can easily compute $<2z\frac{d}{dz}>$ by integration by parts 
(given real wavefunctions).
\begin{eqnarray}
<2z\frac{d}{dz}>=\int\psi2z\frac{d}{dz}\psi=I_{T} \nonumber \\
I_{T}=\psi^{2}2z-\int\psi(2\psi+2z\frac{d\psi}{dz} \nonumber \\
I_{T}=0 - 2 - I_{T}\nonumber \\
I_{T}=- 1\nonumber \\
<2z\frac{d}{dz}>= -1
\end{eqnarray}
This yields the simple result first derived by Lipparini and Stringari [3]
\begin{equation}
\frac{1}{2}[EWS(+) + EWS(-)] = \frac{9}{8\pi} \frac{\hbar^{2}}{M} \sum
 <0| t_{z}+\frac{1}{2}-t_{z}|0> 
\end{equation}
For N=Z we have
\begin{equation}
EWS(+) = EWS(-) = \frac{9}{16\pi} A \frac{\hbar^{2}}{M}
\end{equation}
since in this case, $EWS(+) = EWS(-)$.
For the SUM we obtain 
\begin{equation}
SUM =\frac{EWS(rY^{1}_{k}t_{+})}{\hbar\omega}=\frac{9}{16\pi} A 
\frac{\hbar}{M\omega}
\end{equation}
Finally we get 
\begin{equation}
4 \pi SUM \frac{M \omega}{\hbar}= 9 \frac{A}{4}
\end{equation}
This is the quantity given in Tables 1 and 2 of ref [1]. For $A=12$ we get 27 
for this quantity, confirming the results previously obtained. [1]
\bigskip

\noindent 3) EFFECT OF DIFFERENT FREQUENCIES IN THE X,Y,AND Z DIRECTIONS
\bigskip

To take deformation effects further into account we introduce 
different frequencies in the x, y, and z
directions.  It can be shown that we get the correct result by making 
the following replacement in Eq(12).
\begin{equation}
\frac{1}{\hbar\omega} \rightarrow \frac{1}{3}(\frac{1}{\hbar\omega_{x}}+
\frac{1}{\hbar\omega_{y}}+\frac{1}{\hbar\omega_{z}})
\end{equation}

To obtain this result we must not only consider excitations from 
0p to higher shells but also excitations from 0s to 0p. Note that if the above expression (14)
is expanded in terms of a deformation parameter $\delta$ there will be no linear terms.  

To get an estimate of the size of this effect we use the self consistency 
conditions
\begin{equation}
 \Sigma_{x}\omega_{x} = \Sigma_{y}\omega_{y} = \Sigma_{z}\omega_{z}
\end{equation}
where for $^{12}$C in the asymptotic limit
\begin{eqnarray}
 \Sigma_{x}=\Sigma_{y}=10  \nonumber \\
 \Sigma_{z}=6
\end{eqnarray}
We define $\omega_{0}$ by $\omega_{x}\omega_{y}\omega_{z} =  \omega_{0}^{3}$ and
assume 
volume conservation, i.e. keep $\omega_{0}$ constant.  We then find 
$\omega_{x}= 0.8434 \omega_{0}$ and $\omega_{z}= 1.4057 \omega_{0}$.
We find  $\frac{1}{3}(\frac{1}{\hbar\omega_{x}}+
\frac{1}{\hbar\omega_{y}}+\frac{1}{\hbar\omega_{z}}) =
 \frac{1.0275}{\hbar \omega_{0}}$
 
 There is a very small change in the overall strength.  However $\frac{2}{3}$ of the strength
 is shifted down to $0.8434 \hbar \omega_{0}$ and $\frac{1}{3}$ is
 shifted up to $1.4057 \hbar \omega_{0}$.  (Obviously the energy weighted strength
 does not change in this model.)
\bigskip

\noindent 4)  SPIN DIPOLE AND ORBITAL DIPOLE MODES
\bigskip

We next consider the spin-dipole mode and consider the EWSR in which only the kinetic energy
is taken into account
\begin{eqnarray}
EWS(spin \, multipole) =  \sum_{\lambda M} \sum_{i} \nonumber \\  
\frac{1}{2}<[[Y^{L}(i)s(i)]^{\lambda \dagger}_{M},
[\frac{p^{2}(i)}{2m},[Y^{L}(i)s(i)]^{\lambda}_{M}]> \\
= \sum_{i} \sum_{L,M,M_{L},M_{S},M^{'}_{L},M^{'}_{S}} (L1M_{L}M_{S}| \lambda M)(L1M^{'}_{L}M^{'}_{S}| \lambda M)
 \nonumber \\ 
 \frac{1}{2} 
<[Y^{L \dagger}_{M_{L}}(i)s^{\dagger}_{M_{S}}(i),[\frac{p^{2}(i)}{2m},Y^{L}_{M^{'}_{L}}(i)s_{M^{'}_{S}}(i)>
\end{eqnarray}

Now since

\begin{equation}
\sum_{M,L} (L1M_{L}M_{S}| \lambda M)(L1M^{'}_{L}M^{'}_{S}| \lambda M) = 
\delta_{M_{L},M_{L}^{'}} \delta_{M_{s},M_{s}^{'}}
\end{equation}

We obtain

\begin{eqnarray}
EWS(spin \, multipole)= \frac{1}{2} \sum_{i} \sum_{M_{L}} 
<[Y^{L *}_{M_{L}}(i),[\frac{p^{2}(i)}{2m},
Y^{L}_{M_{L}}(i)]] \sum s^{\dagger}_{M_{s}}s_{M_{s}}> \\
= \sum_{i} EWS(ordinary \, multipole) s(i) \cdot s(i)
\end{eqnarray}

Now $s(i) \cdot s(i)$ is equal to $3/4$ for spin $1/2$ particles.  Hence 
$EWS$(spin multipole) $=$ 3/4 $EWS$(ordinary multipole)
For the spin dipole case in which there is only one excitation energy 
 ($\hbar \omega$) the above relation also holds for summed strength. 
 This was mentioned but not proven by Auerbach and Zamick[1].

As noted in their work and in Table 1 of the present work, the value of SUM for the spin-dipole 
case is 20.25 which is indeed 3/4 of the ordinary dipole 27.

For the orbital dipole case we replace $s(i) \cdot s(i)$ by $l(i) \cdot l(i)$.  This differs
from the spin-dipole case in the sense that $l(i) \cdot l(i)$ is 
state dependent with eigenvalue l(l+1)
.  The value of SUM in Table 1 (summed also over $\lambda$) is 48, all coming from the 
0p shell.
The value of SUM for the ordinary dipole coming from the 0p shell is 27-3=24.  
The factor of 2 is due to the fact that $l(l+1)$ equals two in the 0p shell.

\bigskip

This work was supported by the U.S. Department of Energy under grant
DE-FG02-95ER-40940 and DE-FG02-96ER-40987.  One of us (N.A.) is grateful for the
hospitality afforded to him at Rutgers University.

\newpage
\begin{center}Table 1 \ \ \  Total isovector dipole strength in
$^{12}C(\frac{4\pi m\omega}{\hbar} SUM(\lambda)0^{+} \rightarrow \lambda)$ \\
in parenthesis are the strengths due to excitations from $0s$.
\end{center}

\begin{tabbing}
xxxxxxxxxxxxxxxx\= xxxxxxxxxxxxxxxxxxx\= xxxxxxxxxxxxx\= xxxxxxx \kill

\underline{Dipole} \> \underline{$rY^{1}_{K}t$} \> \> \\
\> Asymptotic \> Spherical \>  \\
$\lambda$ \ \ \> \> \> \\
1 \ \   \> \ \ \ \ \ \ \ 27 \ \ \  \>  \ \ 
\ 27 \ \ \  \> \bigskip \\ 

\\
\\
\\

\underline{Spin Dipole} \> \underline{$r[Y^{1}s]^{\lambda}t$} \> \> \\
\ \ $\lambda$ \ \ \ \> \> \> \\
\ \ 0 \ \ \ \> \ \ \ \ 2.25 (0.25) \> \ \ \ 3.25 (0.75) \> \\
\ \ 1 \ \ \ \> \ \ \ \ 6.75 (0.75) \> \ \ \ 8.25 (1.50) \> \\
\ \ 2 \ \ \ \> \ \underline{\ \ 11.25 (1.25) \ \ } \> \ \underline{ \ 8.75 (0)
\ } \>
\bigskip \\ 
\ \ Sum  \> \ \ \ 20.25 (2.25) \> \ 20.25 (2.25) \> \\
\\
\\
\\
\underline{Orbital Dipole} \> \underline{Dipole $r[Y^{1}\ell]^{\lambda}t$} \> \> \\
$\lambda$ \ \> \> \> \\
0 \ \> \ \ \ \ \ \ 0 \ (0) \> \ \ \ 0 \ (0) \> \\
1 \ \> \ \ \ \ \ \ 14 (0) \> \ \ \ 14 (0) \> \\
2 \ \> \ \ \ \ \underline{ \ 34 (0) \ \ } \> \ \underline{\ \ 34 (0) \ \ }
\> \bigskip \\ 
Sum \> \ \ \ \ \ 48 (0) \> \ \ 48 (0) \> \\

\end{tabbing}

\newpage
\centerline{References}
\bigskip

\begin{enumerate}
\item L. Zamick and N. Auerbach, Nuclear Physics A \underline{658}, 285 (1999)

\item A. Bohr and B. Mottelson, Nuclear Structure, Vol. 1 and Vol. 2
(1969) and (1975), (Benjamin, New York, 1975).

\item W. Lipparini and S. Stringari, Phys. Rep. \underline{175}, 103 (1989)

\end{enumerate}

\end{document}